\documentclass[11pt]{article}
\newcommand\be{\begin{equation}}
\newcommand\ee{\end{equation}}
\newcommand{\fa}{\begin{eqnarray}}
\newcommand{\ffa}{\end{eqnarray}}
\usepackage[dvips]{graphicx}

\def\cc{\circ}
\def\pa{\partial}

\def\Si{\Sigma}
\def\G{\Gamma}
\def\D{\Delta}
\def\O{\omega}
\def\th{\theta}
\def\k{\kappa}
\def\t{\tau}

\def\b{\beta}

\def\e{\epsilon}

\def\a{\alpha}
\def\d{\delta}
\def\l{\lambda}

\begin{document}

\title{The $\hat{Q}$ operator for canonical quantum gravity}
\author{Yongge Ma$^1$\thanks{E-mail: ygma@fis.uncor.edu}
and Yi Ling${}^{2}$\thanks{E-mail: ling@phys.psu.edu}\\
{ }\\
\centerline{\it ${}^1$FaMAF, Universidad Nacional de Cordoba,}\\
\centerline{\it 5000, Cordoba, Argentina}\\
{ }\\ \centerline{\it
${}^2$Center for Gravitational Physics and Geometry}\\
\centerline{\it Department of Physics}\\ \centerline {\it The
Pennsylvania State University}\\ \centerline{\it University Park,
PA, USA 16802}}

\maketitle

\begin{abstract}
\baselineskip=20pt

We study the properties of $\hat{Q}[\O]$ operator on the kinematical
Hilbert space ${\cal H}$ for canonical quantum gravity. Its complete spectrum
with respect to the spin network basis is obtained. It turns out that
$\hat{Q}[\O]$ is diagonalized in this basis, and it is a well defined
self-adjoint operator on ${\cal H}$. The same conclusions are also tenable on
the $SU(2)$ gauge invariant Hilbert space with the gauge invariant spin
network basis.

\end{abstract}

\newpage

\section{Introduction}
\label{sec:1} \baselineskip=20pt

It is well known that considerable progresses have been made in
non-perturbative canonical quantum gravity in the light of
Ashtekar's variables\cite{as86} and loop variables\cite{RS90}. One
of the most remarkable physical results is the evidence for a
quantum discreteness of space at the Planck scale. This is due to
the fact that certain operators corresponding to the measurement
of area and volume have discrete spectra
\cite{RS95,AL97,DR96,AL98}. Also, the kinematical Hilbert space,
${\cal H}$, of the theory has been rigorously defined by
completing the space of all finite linear combinations of
cylindrical functions in the norm induced by Haar measure
\cite{AL95,ba96}.

There is another geometrical operator $\hat{Q}[\O]$ proposed in previous
literatures \cite{ARS92,sm92}, which corresponds to the integrated norm of
any smooth one form $\O_a$ on the 3-manifold. While the operators of area and
volume have been shown to be well defined self-adjoint operators on
${\cal H}$, the general properties of $\hat{Q}[\O]$ are still unclear. We even
do not know if it is well defined on ${\cal H}$. The obstacles are due to two
facts: First, the result of $\hat{Q}[\O]$ operating on a cylindrical function
will involve integrals over edges of the graph on which the function defined,
and hence it is no more a cylindrical function in general; Second,
the current effective regularization technique of smearing the triads in
2-dimension\cite{AL97,GR99} could not be directly applied to the
regularization of $\hat{Q}[\O]$, whose classical expression involves the
square of the triads while there is an integral over 3-dimensional manifold
at last.

As $\hat{Q}[\O]$ operator is rather convenient for constructing
certain weave states in the study of the classical approximation of the
quantum theory \cite{ARS92,ML00}, the present paper is devoted to study the
properties of $\hat{Q}[\O]$ on ${\cal H}$ in order to lay a foundation of its
applications. To bypass the above mentioned obstacles, we will use a
3-dimensional smearing
function for regularization. Then, instead of acting the regulated
$\hat{Q}[\O]$ on general cylindrical functions, we will operate it on spin
network states which form a complete orthonormal basis in ${\cal H}$. It turns
out that the operation gives a real discrete spectrum, which is in the same
form as its eigenvalues on coloured loop states. Thus, $\hat{Q}[\O]$
is a well defined symmetric operator in ${\cal H}$. A further discussion shows
it is also self-adjoint.

We work in the real Ashtekar formalism defined over an oriented 3-manifold
$\Si$ \cite{ba95}. The basic variables are real $SU(2)$ connections, $A_a^i$,
as the configuration and the densitized(weight 1) triads, $E^b_j$,
corresponding to the conjugate momentum. We use $a,b,\cdots$ for spatial
indices and $i,j,\cdots$ for internal $SU(2)$ indices. The basic variables
satisfy
\be\label{0}
\{A_a^i(x), E^b_j(y)\}=G\d^i_j\d^b_a\d^3(x,y),
\ee
where $G$ is the usual gravitational constant.

\section{$\hat{Q}$ operator and its regularization}
\label{sec:2}

The operator $\hat{Q}[\O]$ is  constructed to represent the classical quantity
\cite{ARS92,sm92}
\be \label{1}
Q[\O]= \int d^3x \sqrt{E^a_i(x)\O_a(x)E^{bi}(x)\O_b(x)},
\ee
where $\O_a$ is any smooth
1-form on $\Si$ which makes the integral meaningful and the integral is
well defined since the integrand is a density of weight 1.
If we know $Q[\O]$ for all smooth $\O_a$, the triad $E^a_i$ can be
reconstructed up to local $SU(2)$ gauge transformations. Hence, the collection
of $Q[\O]$ provides a good coordinates system on the space of the
triads fields.

Since $E^a_i$ represents the conjugate momentum of the configuration variable
$A_a^i$, the formal expression of the corresponding momentum operator would be
some functional derivative with respect to $A_a^i$, i.e.,
\be \label{2}
\hat{E}^a_i(x)=-iG\hbar {\d\over \d A_a^i(x)}.
\ee
This is an operator-valued distribution rather than a genuine operator, hence
it has to be integrated against smearing functions in order to be well
defined. Our aim is to construct a well defined operator $\hat{Q}[\O]$
corresponding to the classical quantity $Q[\O]$. We begin with a formal
expression obtained by replacing $E^a_i$ in Eq.(\ref{1}) by the
operator-valued distribution $\hat{E}^a_i$, and then regulate it by
3-dimensional smearing functions.

Let $f_\e(x,y)$ be a 1-parameter family of fields on $\Si$ which tends to
$\d(x,y)$ as $\e$ tends to zero, such that $f_\e(x,y)$ is a density of weight
1 in $x$ and a function in $y$. We then define the smeared version of
$E^a_i(x)\O_a(x)$ as:
\be \label{3}
[E_i\O]_f(x):=\int d^3y f_\e(x,y)E^a_i(y)\O_a(y).
\ee
Hence, $[E_i\O]_f(x)$ tends to $E^a_i(x)\O_a(x)$ as $\e$ tends to zero. Then
$Q[\O]$ can be regulated as:
\be \label{4}
Q[\O]=\lim_{\e\rightarrow 0} \int d^3x\left([E_i\O]_f(x)[E^i\O]_f(x)
\right)^{1\over 2}.
\ee
To go over to the quantum theory, we simply replace $E^a_i$ by $\hat{E}^a_i$
and obtain
\be \label{5}
\hat{Q}[\O]=\lim_{\e\rightarrow 0} \int d^3x\left([\hat{E}_i\O]_f(x)
[\hat{E}^i\O]_f(x)\right)^{1\over 2},
\ee
where
\be \label{6}
[\hat{E}_i\O]_f(x):=\int d^3y f_\e(x,y)\hat{E}^a_i(y)\O_a(y)
=-iG\hbar\int d^3y f_\e(x,y)\O_a(y)\left({\d\over \d A_a^i(y)}\right).
\ee
By operating the regulated $\hat{Q}[\O]$ on spin network states in next
section, we will show that it is a well defined symmetric operator in the
kinematical Hilbert space and admits self-adjoint extensions. For technical
reasons, we attach the following concreteness to the smearing function $f_\e$
for sufficiently small $\e >0$: (i) $f_\e(x,y)$ is non-negative; and (ii) for
any given $y$, $f_\e(x,y)$ has compact support in $x$ which is a 3-dimensional
box, $U_{\e}$, of coordinate height $\e^\b$, $1<\b<2$, and square horizonal
section, $S_{\e}$, of coordinate side $\e$, and with $y$ as its centre. These
conditions are in the same spirit as that in Ref.\cite{FLR97}. More
concretely, $f_\e(x,y)$ can be constructed as follows. Take any 1-dimensional
non-negative function $\th(x)$ of compact support $[-{1\over 2}, {1\over 2}]$
on $R$ such that $\int dx \th(x)=1$, and set
\be \label{7}
f_\e(x,y)=\left({1\over \e^{2+\b}}\right)\th\left({x_1-y_1\over \e}\right)
\th\left({x_2-y_2\over \e}\right)\th\left({x_3-y_3\over \e^\b}\right).
\ee

\section{Action of $\hat{Q}$ on spin network basis}
\label{sec:3}

\subsection{Preliminaries}
\label{sec:3.1}
It has been shown that spin networks play a key role in non-perturbative
quantum gravity \cite{RS95b,ba96,AL97}. Consider a graph $\G$ with $n$ edges
$e_I$, $I=1,\cdots,n$, and $m$ vertices $v_\a$, $\a=1,\cdots,m$, embedded in
the 3-manifold $\Si$. To each $e_I$ we assign a non-trivial irreducible spin
$j_I$ representation of $SU(2)$. This is called a colouring of the edge. Next,
consider a vertex $v_\a$, say a $K$-valent one, i.e., there are $K$ edges
$e_1,\cdots,e_K$ meeting at $v_\a$. Let ${\cal H}_{j_1},\cdots,{\cal H}_{j_K}$
be the Hilbert spaces of the representations, $j_1,\cdots,j_K$, associated to
the $K$ edges. Consider the tensor product of these spaces
${\cal H}_{v_\a}={\cal H}_{j_1}\otimes\cdots\otimes{\cal H}_{j_K}$, and fix,
once and for all, an orthonormal basis, $N_\a$, in ${\cal H}_{v_\a}$. This is
called a colouring of the vertex. A (non-gauge invariant) spin network, $S$,
is then defined as the embedded graph whose edges and vertices have been
coloured.

The holonomy of the $SU(2)$ connection $A_a^i$ along any edge $e_I$
is an element of $SU(2)$ and can be expressed as:
\be\label{a1}
h[A,e_I]={\cal P}exp\int_{e_I}ds\dot{e}_I(s)A_a^i(e_I(s))\t_i,
\ee
where ${\cal P}$ denotes path ordering and $\t_i$ are the $SU(2)$ generators
in the fundamental representation. The (non-gauge invariant) spin network
state, $\Psi_S(A)$, based on $S$ is defined as:
\be\label{a2}
\Psi_S(A)=\bigotimes_{e_I\in\G}j_I\left(h[e_I]\right)\cdot
\bigotimes_{v_\a\in\G}
N_\a,
\ee
where $j_I\left(h[e_I]\right)$ is the representation matrix of the holonomy
$h[e_I]$ in the spin $j_I$ representation associated to the edge $e_I$, and
the holonomy matrices are constructed with the vector $N_\a$ at each vertex
$v_\a$ where the edges meet. By varying the graph, the colours of the edges,
and the colours of the vertices, we obtain a family of spin network states. It
turns out that these states form a complete orthonormal basis in the
kinematical Hilbert space ${\cal H}$ \cite{AL97,GR99}.

Since a $SU(2)$ gauge transformation acts on a spin network state simply by
$SU(2)$ transforming the colouring of the vertices $N_\a$, it is easy to
recover the gauge invariant\footnote{The gauge invariance discussed in this 
paper is restricted to that of internal $SU(2)$, while the whole gauge 
invariance of a gravitational theory should also involve that of 
4-dimensional diffeomorphism.} spin network states by colouring each vertex 
with a $SU(2)$ invariant basis. These states form a complete orthonormal 
basis in
the $SU(2)$ gauge invariant Hilbert space ${\cal H}_0$ \cite{RS95b,ba96,AL95}.

It is obvious from Eqs. (\ref{2}) and (\ref{a1}) that the action of
$\hat{E}^a_i(x)$ on a holonomy $h[e_I]$ yields
\be\label{a3}
\hat{E}^a_i(x)\cc h[e_I]=-il_p^2\int_{e_I}ds\dot{e}_I^a(s)\d^3(x,e_I(s))
h_I[1,s]\t_ih_I[s,0],
\ee
where $l_p=\sqrt{G\hbar}$ is the Planck length.

\subsection{Spectrum of $\hat{Q}$}
\label{sec:3.2}

We first apply the operator $[\hat{E}_i\O]_f(x)$ defined by Eq.(\ref{6}) to
the spin network state $\Psi_S$,
\fa \label{7a}
[\hat{E}_i\O]_f(x)\cc\Psi_S(A)&=&-il_p^2\sum_{I=1}^n\int d^3y f_\e(x,y)\O_a(y)
\left[{\d\over \d A_a^i(y)}j_I(h_I)_{lm}\right]
\left({\partial\Psi_S\over\partial j_I(h_I)_{lm}}\right) \nonumber\\
&=&-il_p^2\sum_{I=1}^n\int d^3y f_\e(x,y)\int_{e_I}dt\dot{e}_I^a(t)\O_a(y)
\d^3(y,e_I(t))j_I\left(h_I[1,t]\t_ih_I[t,0]\right)_{lm}
\left({\partial\Psi_S\over\partial j_I(h_I)_{lm}}\right) \nonumber\\
&=&-il_p^2\sum_{I=1}^n\int_{e_I}dt\dot{e}_I^a(t)\O_a(e_I(t))f_{\e}(x,e_I(t))
Tr\left[j_I\left(h_I[1,t]\t_ih_I[t,0]{\partial\over\partial h_I}\right)\right]
\cc\Psi_S(A),
\ffa
where $l$ and $m$ are indices in ${\cal H}_{j_I}$ associated to $e_I$.
Repeating the action of $[\hat{E}^i\O]_f(x)$ on Eq.(\ref{7a}), we have
\fa \label{8}
[\hat{E}^i\O]_f(x)[\hat{E}_i\O]_f(x)\cc\Psi_S&=&-l_p^4\sum_{I=1}^n
\int_{e_I}dt\dot{e}_I^a(t)\O_a(e_I(t))\sum_{J=1}^n\int_{e_J}ds\dot{e}_J^b(s)
\O_b(e_J(s))f_{\e}(x,e_I(t))f_{\e}(x,e_J(s))\nonumber\\
&{ }&Tr\left[j_J\left(h_J[1,s]\t^ih_J[s,0]{\pa\over\pa h_J}\right)\right]
Tr\left[j_I\left(h_I[1,t]\t_ih_I[t,0]{\pa\over\pa h_I}\right)\right]\cc\Psi_S
\nonumber\\
&{ }&-l_p^4\sum_{I=1}^n\int_{e_I}dt\dot{e}_I^a(t)\O_a(e_I(t))
f_{\e}(x,e_I(t))
\left(\int_t^1ds\dot{e}_I^b(s)\O_b(e_I(s))f_{\e}(x,e_I(s))\right.\nonumber\\
&{ }&Tr\left[j_I\left(h_I[1,s]\t^ih_I[s,t]\t_ih_I[t,0]{\pa\over\pa h_I}
\right)\right]
+\int_0^t ds\dot{e}_I^b(s)\O_b(e_I(s))f_{\e}(x,e_I(s))\nonumber\\
&{ }&\left.Tr\left[j_I\left(h_I[1,t]\t_ih_I[t,s]\t^ih_I[s,0]{\pa\over\pa h_I}
\right)\right]\right)\cc\Psi_S.
\ffa
We denote respectively the first and second terms in the right hand side of
Eq.(\ref{8}) as $A$ and $B$. Consider first the term $A$. Note that a spin
network state can always be written as:
\be \label{9}
\Psi_S(A)=j_I\left(h[e_I]\right)_{lm}\Psi_{S-e_I}^{lm}(A),
\ee
where $\Psi_{S-e_I}^{lm}(A)={\pa\Psi_S\over\pa j_I(h_I)_{lm}}$ is independent
of $h[e_I]$. Hence, we can choose $\e$ sufficiently small, such that the term
$A$ vanishes unless the support $U_{\e}$ of the smearing function $f_{\e}$
contains a vertex $v_{\a}$ of the spin network as its centre. It then turns
out
\fa \label{10}
A&=&-l_p^4\sum_{I,J=1}^n
\int_{e_I}dt\dot{e}_I^a(t)\O_a(e_I(t))\int_{e_J}ds\dot{e}_J^b(s)
\O_b(e_J(s))[f_{\e}(x,v_{IJ})]^2 \nonumber\\
&{ }&Tr\left[j_J\left(h_J[1,s]\t_ih_J[s,0]{\pa\over\pa h_J}\right)\right]
Tr\left[j_I\left(h_I[1,t]\t_ih_I[t,0]{\pa\over\pa h_I}\right)\right]\cc\Psi_S
\nonumber\\
&=&-l_p^2\sum_{\a=1}^m[f_{\e}(x,v_{\a})]^2\sum_{I_\a,J_\a}
\int_{e_I}dt\dot{e}_I^a(t)\O_a(e_I(t))\int_{e_J}ds\dot{e}_J^b(s)
\O_b(e_J(s)) \nonumber\\
&{ }&Tr\left[j_J\left(h_J[1,s]\t_ih_J[s,0]{\pa\over\pa h_J}\right)\right]
Tr\left[j_I\left(h_I[1,t]\t_ih_I[t,0]{\pa\over\pa h_I}\right)\right]\cc\Psi_S.
\ffa
To simplify technicalities, given a 1-form $\O_a$ we choose $f_{\e}$ such
that at each vertex, $\O_a$ is a normal covector of the horizonal section
$S_\e$ of $U_\e$, i.e., $\O_a(v_\a)=|\O(v_\a)|(dx_3)_a$. (Note that the 
special $f_{\e}$ is chosen here in order to obtain a succinct expression of 
$A$, see Eq.(\ref{15}), while the final result that $A$ makes no contribution 
to the spectrum of $\hat{Q}$ is independent of this choice.) Since $\e^\b$ goes 
to
zero faster than $\e$, for sufficiently small $\e$ the
edge $e_I$ which meets the vertex $v_\a$ would cross the top or bottom of the
box $U_\e$ if it is not tangent to $S_\e$ at $v_\a$. Also, it follows from
$\b<2$ that any edge tangential to $S_\e$ at $v_\a$ exits $U_\e$ from the
side, irrespectively from its second (and higher) derivatives, and gives a
vanishing contribution to $A$ as $\e$ goes to zero. Thus, if we first
consider only the ``outgoing'' edges from the vertices, it turns out
\be \label{11}
\int_{e_I}dt\dot{e}_I^a(t)\O_a(e_I(t))f_{\e}(x,v_{\a})={1\over 2}\k_I\e^\b
|\O(v_\a)|f_\e(x,v_\a)+O(\e),
\ee
where
\fa \label{12}
\k_I:=\left\{
\begin{array}{lll}
0, & \mbox{ if $e_I$ is tangent to $S_\e$} \\
1, &\mbox{ if $e_I$ lies above $S_\e$} \\
-1,&\mbox{ if $e_I$ lies below $S_\e$}
\end{array}
\right.
\ffa
and hence Eq.(\ref{10}) becomes
\be \label{13}
A_{out}=-{1\over 4}l_p^4\sum_{\a=1}^m[f_{\e}(x,v_{\a})\e^\b|\O(v_\a)|]^2
\sum_{I_\a,J_\a}[\k_I\k_J L_I^i L_J^i+O(\e)]\cc\Psi_S,
\ee
where
\be\label{14}
L_I^i\cc\Psi_S:=Tr\left[j_I\left(h[e_I]\t^i{\pa\over\pa h[e_I]}\right)\right]
\cc\Psi_S.
\ee
Including the ``incoming'' edges to the vertices, the final expression of $A$
reads
\be\label{15}
A=-{1\over 4}l_p^4\e^{2\b}\sum_{\a=1}^m[f_{\e}(x,v_{\a})|\O(v_\a)|]^2
\sum_{I_\a,J_\a}[\k_I\k_J X_I^i X_J^i+O(\e)]\cc\Psi_S,
\ee
where
\fa\label{16}
X_I^i\cc\Psi_S:=\left\{
\begin{array}{ll}
Tr\left[j_I\left(h[e_I]\t^i{\pa\over\pa h[e_I]}\right)\right]\cc\Psi_S,
&\mbox{if $e_I$ is outgoing}\\
Tr\left[j_I\left(-\t^i h[e_I]{\pa\over\pa h[e_I]}\right)\right]\cc\Psi_S,
&\mbox{if $e_I$ is incoming.}
\end{array}\right.
\ffa
Note that $\D_{S_\e,v_\a}=\sum_{I_\a,J_\a}\k_I\k_J X_I^i X_J^i$ is the vertex
operator associated with $S_\e$ and $v_\a$ in arbitrary spin representations,
which has been fully investigated \cite{AL97,AL95b}. A discussion similar to
that in Ref.\cite{AL97} leads that the spin network state $\Psi_S$ is an
eigenvector of $-\D_{S_\e,v_\a}$ with eigenvalue:
\be\label{17}
\l_{S_\e,v_\a}=2j^{(d)}(j^{(d)}+1)+2j^{(u)}(j^{(u)}+1)-j^{(d+u)}(j^{(d+u)}+1),
\ee
where $j^{(d)}$, $j^{(u)}$ and $j^{(d+u)}$ are half integers subject to the
condition \\ $j^{(d+u)}\in \{|j^{(d)}-j^{(u)}|,|j^{(d)}-j^{(u)}|+1,\cdots,
j^{(d)}+j^{(u)}\}$.

Now consider the second term $B$. For small $\e$,
$f_{\e}(x,e_I(t))f_{\e}(x,e_I(s))$ is non-zero only for the parameters
satisfying $t=s+O(\e)$, where we have
\fa\label{19}
Tr\left[j_I\left(h_I[1,t]\t_ih_I[t,s]\t^ih_I[s,0]{\pa\over\pa h_I}
\right)\right]\cc\Psi_S &=&
\left(Tr\left[j_I\left(h_I[1,t]\t_i\t^ih_I[t,0]{\pa\over\pa h_I}\right)\right]
+O(\e)\right)\cc\Psi_S
\nonumber\\
&=&\left(-j_I(j_I+1)Tr\left[j_I\left(h_I[e_I]{\pa\over\pa h[e_I]}\right)\right]
+O(\e)\right)\cc\Psi_S \nonumber\\
&=&-[j_I(j_I+1)+O(\e)]\Psi_S,
\ffa
here $-j_I(\t_i\t^i)=j_I(j_I+1)$ is the Casimir operator of $SU(2)$.
Substituting Eq.(\ref{19}) into $B$, we obtain
\be\label{20}
B=l_p^4\sum_{I=1}^n\left[\int_{e_I}dt\dot{e}_I^a(t)\O_a(e_I(t))
f_{\e}(x,e_I(t))\right]^2[j_I(j_I+1)+O(\e)]\Psi_S.
\ee
It then follows from Eqs. (\ref{15}) and (\ref{20}) that
\fa\label{21}
[\hat{E}^i\O]_f(x)[\hat{E}_i\O]_f(x)\cc\Psi_S={1\over 4}\e^{2\b}l_p^4
\sum_{\a=1}^m[f_{\e}(x,v_{\a})|\O(v_\a)|]^2(\l_{S_\e,v_\a}+O(\e))\Psi_S
\nonumber\\
+l_p^4\sum_{I=1}^n\left[\int_{e_I}dt\dot{e}_I^a(t)\O_a(e_I(t))
f_{\e}(x,e_I(t))\right]^2[j_I(j_I+1)+O(\e)]\Psi_S,
\ffa
which implies that $[\hat{E}^i\O]_f(x)[\hat{E}_i\O]_f(x)$ is a well defined
non-negative operator and hence has a well defined square-root. Since we have
chosen $\e$ to be sufficiently small, for any given $x\in\Si$,
$f_{\e}(x,v_{\a})$ is non-zero for at most one vertex, and $f_{\e}(x,e_I(t))$
is non-zero for at most a piece of one edge where its vertices are not
included. Therefore we can take the sum over $v_\a$ and $e_I$ out side the
square root and obtain
\fa\label{22}
\left([\hat{E}^i\O]_f(x)[\hat{E}_i\O]_f(x)\right)^{1\over 2}\cc\Psi_S
={1\over 2}\e^{\b}l_p^2\sum_{\a=1}^mf_{\e}(x,v_{\a})|\O(v_\a)|
(\l_{S_\e,v_\a}+O(\e))^{1\over 2}\Psi_S \nonumber\\
+l_p^2\sum_{I=1}^n\left|\int_{e_I}dt\dot{e}_I^a(t)\O_a(e_I(t))
f_{\e}(x,e_I(t))\right|[j_I(j_I+1)+O(\e)]^{1\over 2}\Psi_S.
\ffa
Now we can remove the regulator. Taking the limit $\e\rightarrow 0$ and
integrating over $\Si$, the first term in the right hand side of Eq.(\ref{22})
vanishes due to the factor $\e^\b$. We thus conclude that the action of
$\hat{Q}[\O]$ on spin network states yields
\be\label{23}
\hat{Q}[\O]\cc\Psi_S(A)=l_p^2\sum_{I=1}^n\left[\int_{e_I}dt
|\dot{e}_I^a(t)\O_a(e_I(t))|\sqrt{j_I(j_I+1)}\right]\Psi_S(A).
\ee
Therefore, spin network states are also eigenvectors of $\hat{Q}[\O]$. The
complete spectrum of $\hat{Q}[\O]$ with respect to the spin network basis in
the Hilbert space ${\cal H}$ is obtained.

\section{Discussions}
\label{sec:4}

The general properties of $\hat{Q}[\O]$ operator are implied by Eq.(\ref{23}).
In contrast to the volume operator which acts only on vertices
\cite{DR96,AL98}, $\hat{Q}[\O]$ acts only on edges of spin networks. Hence,
the spin network states based on a same graph with same colouring of the edges
are all degenerate with respect to this operator. As a result, the action of
$\hat{Q}[\O]$ on the gauge invariant spin network states gives the same result
as Eq.(\ref{23}). In this sense, the spectrum of $\hat{Q}[\O]$ respects the
physically relevant states in ${\cal H}$.

There are alternative approaches to regulate $\hat{Q}[\O]$ and calculate its
spectrum. One could also apply the blocking regularization technique of
Ref.\cite{FLR97}, then express $\hat{Q}[\O]$ by the loop operator
${\cal T}^{ab}$ up to $O(\e)$, whose action on spin network states is obtained
from the recoupling theory \cite{DR96}. It is not difficult to check that
this approach will give the same result as we have obtained. By restricting
the support of the regulator, our approach reveals the inherent relation of
the two approaches.

We have shown that $\hat{Q}[\O]$ is diagonalized in the spin network basis
with real eigenvalues, hence it is a well defined symmetric operator in the
kinematical Hilbert space ${\cal H}$. Moreover, it is obvious from Eqs.
(\ref{5}) and (\ref{6}) that the expression of $\hat{Q}[\O]$ is purely real,
and hence it commutes with the complex conjugation. Therefore, it follows from
Von Neumann's theorem\cite{RS2} that $\hat{Q}[\O]$ admits self-adjoint
extensions on ${\cal H}$. The same reasons lead that $\hat{Q}[\O]$ is also
self-adjoint on the gauge invariant Hilbert space ${\cal H}_0$.

The discrete spectrum of $\hat{Q}[\O]$ shows a quantum discreteness of the
space at the Planck scale, corresponding to the measurement of the integrated
norm of any smooth one forms.

\subsection*{Acknowledgements}
Y. Ma would like to thank discussions with Profs. C. N. Kozameh, O. M. 
Moreschi, and especially O. A. Reula for his enlightening comments and 
suggestions, and acknowledge support from FONCYT BID 802/OC-AR PICT: 00223. 
Y. Ling is grateful to Abhay Ashtekar for discussion, and was supported by 
the NSF through grant PHY95-14240, a gift from the Jesse Phillips Foundation, 
and a Braddock fellowship from the Department of Physics at Pennsylvania
State University.

\end{document}